\documentclass[apjl,twocolumn]{openjournal}
\usepackage{amsmath}
\usepackage{booktabs}
\usepackage{multirow}
\usepackage{color}
\usepackage{soul}
\usepackage{upgreek}

\usepackage[dvipsnames]{xcolor} 

\usepackage[breaklinks,colorlinks,citecolor=blue,urlcolor=blue]{hyperref}

\renewcommand{\arraystretch}{1.2}  
\usepackage{listings}
\usepackage{color}
\definecolor{dkgreen}{rgb}{0,0.6,0}
\definecolor{gray}{rgb}{0.5,0.5,0.5}
\definecolor{mauve}{rgb}{0.58,0,0.82}
\definecolor{golden}{rgb}{0.86,0.65,0.01}
\defcitealias{masunaga1998}{MMI}
\defcitealias{lowlyndenbell1976}{LLB}
\defcitealias{masunaga:opacity.limit.tau}{MI99}
\begin{document}

\title{The opacity limit\vspace{-1.5cm}} 

\author{Michael Y. Grudi\'{c}$^{1,\dagger}$}
\author{Philip F. Hopkins$^{2}$}

\affiliation{$^1$Carnegie Observatories, 813 Santa Barbara St, Pasadena, CA 91101, USA}
\affiliation{$^{2}$TAPIR, Mailcode 350-17, California Institute of Technology, Pasadena, CA 91125, USA}

\thanks{$\dagger$NASA Hubble Fellow}
\email{Corresponding author: mgrudic@carnegiescience.edu}

\begin{abstract}
The opacity limit is an important concept in star formation: isothermal collapse cannot proceed without limit, because eventually cooling radiation is trapped and the temperature rises quasi-adiabatically, setting a minimum Jeans mass $M_{\rm J}^{\rm min}$. Various works have considered this scenario and derived expressions for $M_{\rm J}^{\rm min}$, generally $\sim 10^{-3}-10^{-2}M_\odot$ in normal star-forming conditions, but with conflicting results about the scaling with ambient conditions and material properties. We derive expressions for the thermal evolution of dust-cooled collapsing gas clumps in various limiting cases, given a general ambient radiation field ($u_{\rm rad}$, $T_{\rm rad}$) and a general power-law dust opacity law $\sigma_{\rm d} = A_{\rm d} T^{\beta}$. By accounting for temperature evolution self-consistently we rule out a previously-proposed regime in which the adiabatic transition occurs while the core is still optically-thin. If the radiation field is weak or dust opacity is small, $M_{\rm J}^{\rm min}$ is insensitive to dust properties/abundance ($\sim A_{\rm d}^{-\frac{1}{11}}-A_{\rm d}^{-\frac{1}{15}}$), but if the radiation field is strong and dust is abundant it scales $\propto A_{\rm d}^{1/3}$. This could make the IMF less bottom-heavy in dust-rich and/or radiation-dense environments, e.g. galactic centers, starburst galaxies, massive high-$z$ galaxies, and proto-star clusters that are already luminous. 
\end{abstract}

\maketitle  

\section{Introduction}
We live in a hierarchical universe, consisting of nested structures spanning a vast range of scales from the cosmic web down to to moons and asteroids. It is natural for inhabitants of such a universe to ask: where does structure formation stop? Gravity is scale-free but other processes are not, so surely at times the laws of physics must step in and say no more, preventing some structures from spawning substructures \citep{Hoyle_1953}.

This question is particular relevant for understanding star formation and the initial mass function (IMF). Stars in the present-day Universe form by fragmentation and accretion within cold, magnetized, supersonically-turbulent giant molecular clouds (GMCs) that are roughly isothermal over a wide range of densities. The equations of self-gravitating isothermal magnetohydrodynamics are scale-free \citep{mckee_sf_theory}, and hence do not unambiguously imply that GMCs should fragment to objects of stellar mass instead of sub-fragmenting to smaller objects \citep{martel_numerical_sim_convergence, sf_big_problems,guszejnov_isothermal_collapse}. The Jeans mass,
\begin{equation}
    M_{\rm J} = \frac{\uppi^{5/2}}{6}\left(\frac{k_{\rm B}T}{G m_{\rm p} \mu}\right)^{3/2} \rho^{-1/2},
    \label{eq:jeansmass}
\end{equation}
can become arbitrarily small if the density $\rho$ continues to increase at fixed $T$ (assuming $\mu \approx 2.4$ for molecular gas). Why, then, does the observed IMF almost always have a feature around $\sim 0.1-0.3 M_\odot$, with increasing rarity of lower-mass objects? 

One possibility is that radiative transfer in the interstellar medium places restrictions on when collapse and fragmentation can proceed: the opacity limit. \citet{lowlyndenbell1976} (hereafter \citetalias{lowlyndenbell1976}), \citet{rees1976}, and \citet{silk:1977.opacity.limit} solved the balance of heating and cooling in collapsing, opaque, Jeans-mass clumps, and argued that if optically-thick objects can no longer cool, they will evolve quasi-adiabatically and the Jeans mass must rise. When essentially any dust is present, the opacity of dust to its own $\sim 1 \,\rm mm$ thermal emission sets this limit, and the predicted minimum Jeans mass is on the order of $10^{-3}-10^{-2}\,M_\odot$. There is some sensitivity to the precise hydrodynamics and geometry of the collapse \citep{boyd.whitworth:2005.opacity.limit,whitworth:2006.min.mass}, but weak sensitivity to the opacity: for an opacity law $\sigma_{\rm d} = A_{\rm d} T^{\beta}$, $M_{\rm J}^{\rm min} \propto A_{\rm d}^{-\frac{1}{4\beta+7}}$, $\beta \sim 1-2$. This minimum Jeans mass is much smaller than a star, but the dynamics of accretion after hydrostatic core formation may have a multiplicative effect on the typical mass that a fragment eventually accretes  \citep{Lee_Hennebelle_2018_IC,Hennebelle_2019_fhsc_tidal,Colman_Teyssier_2019_tidal_screening}.

\citet{masunaga:opacity.limit.tau} \citepalias[hereafter][]{masunaga:opacity.limit.tau} revisited the opacity limit, and were able to calibrate the uncertain form factors in this theory to a suite of 1D radiation hydrodynamics simulations of protostellar collapse \citep{masunaga1998} \citepalias[hereafter][]{masunaga1998}. They corrected the misconception that the condition of being optically-thick ($\tau \gtrsim 1$) was sufficient to induce adiabatic evolution, showing that adiabatic evolution begins when the compressional heating rate exceeds the maximum radiative energy diffusion rate. Their prediction for $M_{\rm J}^{\rm min}$ is rather more sensitive, with one regime (their ``Case 1") in which $M_{\rm J}^{\rm min} \propto A_{\rm d}^{-1}$, which would make it difficult to make low-mass stars in dust-poor environments such as proto-globular clusters and ultra-faint dwarf galaxies. 

We believe the ``Case 1'' opacity limit of \citetalias{masunaga:opacity.limit.tau} is unrealistic, and are not aware of a work that contains the general set of correct expressions for the opacity-limited density, temperature, and Jeans mass, and related quantities like the first hydrostatic core (``Larson'') mass. For this reason, we revisit the subject of dust-cooled protostellar collapse, writing down approximations for the temperature evolution of collapsing fragments as a function of density in various asymptotic regimes, and arriving at expressions for the opacity-limited Jeans mass and the associated transition density $n_{\rm ad}$ that demarcates the quasi-isothermal and -adiabatic regimes, while correcting some unrealistic assumptions made in previous works. With these expressions in hand we then comment briefly on the scaling of the minimum Jeans mass in different environments, and implications for the IMF.

\section{Model for dust-cooled collapse}
\label{sec:physics}
Consider the scenario of a spherical, self-gravitating, thermally-supported gas clump undergoing approximately self-similar, nearly-isothermal collapse, as has been studied extensively in previous work \citep{larson_1969,penston_1969,masunaga1998,vaytet_2017_protostellar_collapse,2018A&A...618A..95B}. This is a highly-idealized scenario, and other works have considered scenarios that could be more representative of the 3D turbulent collapse process of molecular clouds \citep{boyd.whitworth:2005.opacity.limit,whitworth:2006.min.mass}, but such details only affect the dimensionless prefactors on the opacity-limited quantities of interest, which are themselves degenerate with dust properties. Hence we default to the simple spherical setup, which contains the essential physics, and which has an exhaustive literature of more-detailed calculations that can guide our analytic model.
\subsection{Assumptions}
\label{sec:assumptions}
We solve for the thermal evolution of the collapsing core under the following simplifying assumptions and approximations, all of which follow previous work:
\begin{enumerate}
    \item Dust is the only coolant. This is generally valid at the high ($\gtrsim 10^5 \,\rm cm^{-3}$) densities characteristic of prestellar collapse, where dust emission does tend to exceed molecular and fine-structure emission. A dust-coupled phase should only be absent in extremely metal-poor ($\lesssim 10^{-4} Z_\odot$) or metal-pristine star-forming conditions \citep{sharda.krumholz:2022,klessen.glover:2023.first.stars}.
    \item The dust temperature $T_{\rm d}$ and gas temperature $T$ are equal. The two temperatures should asymptotically approach each other because the rate of dust-gas heat transfer per particle scales $\propto n_{\rm H}\sqrt{T}\left(T-T_{\rm d}\right)$, and no plausible heating mechanism scales as steeply with density if $T$ is nearly isothermal, so $T-T_{\rm d}$ will drop.
    \item The \citetalias{lowlyndenbell1976} assumption that effective shielding length $l=\tau/\sigma n_{\rm H}$ for determining the optical depth is on the order of the Jeans length $\lambda_{\rm J} = \sqrt{\uppi c_{\rm s}^2/\left(G\rho\right)}$:
    \begin{equation}
        l = f_{\rm J} \lambda_{\rm J},
    \end{equation}
    where the numerical prefactor $f_{\rm J} \approx 0.75$ was measured from simulations by \citetalias{masunaga1998}.
    \item The core has no substructure in velocity, gas density, or dust density. This is not so clearly justifiable in nature, and is simply our choice of problem scope. Turbulent/convective motions or clumping of gas and dust could all conceivably affect the rate of energy transport in general and challenge the opacity limit scenario, but this is beyond our scope.
    \item The temperature structure before the adiabatic transition follows from equating the heating and cooling rate. This is only especially accurate when $t_{\rm cool}<<t_{\rm dyn}$, but \citetalias{masunaga1998} showed that it is a fair description even when $t_{\rm cool} \sim t_{\rm dyn}$.
    \item The core is bathed in an isotropic radiation field specified by an energy density $u_{\rm rad}$ and an effective absorption-weighted temperature $T_{\rm rad}$, which need not be in local thermodynamic equilibrium (i.e. $u_\mathrm{ rad}\neq aT_{\rm rad}^4$) in general. This choice is motivated by star-forming conditions in the Solar neighborhood, where typically $T_{\rm rad}\sim 20 \rm K$ for the IR component, and $u_{\rm rad} \sim 1 \rm \,eV\,cm^{-3} << a T_{\rm rad}^4$. This choice of radiation field and the dust opacity constitute our parameter space, and the temperature structure follows from the balance of heating and cooling.
    \item The dust mass fraction, grain size distribution, and composition remain constant throughout the collapse. This may be the least-realistic assumption of current protostellar collapse studies: in dense core conditions many processes can affect the dust grain properties, with a tendency for grains to grow and the grain size distribution to become more top-heavy \citep{1994ApJ...430..713W,2017ApJ...837...78H,2022MNRAS.515.4780T,2023MNRAS.518.3326L}. A transition in the dust properties to a new equilibrium independent of initial conditions is all but inevitable, because the processes regulating grain size typically operate on a collisional timescale $\propto n^{-1}T^{-1/2}$ that will generally overtake the free-fall time $\propto n^{-1/2}$. 
\end{enumerate}

\subsection{Dust model}

Both optically-thin dust emission and absorption depend on the Planck-mean dust opacity, while optically-thick emission and absorption depend on the Rosseland mean opacity. Will will ignore the distinction here because they tend to be close in value, and the key aspect of the problem is how they scale with temperature, which is similar \citep{semenov_2003}. For dust temperatures $\lesssim 100 \rm K$ the dust cross section per H nucleus may be approximated as
\begin{equation}
    \sigma_{\rm d} =  A_{\rm d} T^\beta,
    \label{eq:sigmadust}
\end{equation}
where $T$ is either the grain temperature $T_{\rm d}$ that determines the emission cross-section, or the radiation temperature $T_{\rm rad}$ that determines the absorption cross-section. The normalization $A_{\rm d}$ rolls together all possible variation due to the abundance, composition, and grain size distribution of dust:
\begin{equation}
    A_{\rm d} = \sigma_{\rm 0} \hat{\sigma}_{\rm d} Z_{\rm d} \left(10\,\mathrm{K}\right)^{-\beta},
\end{equation}
where $\sigma_{\rm 0}=4\times 10^{-26}\,\rm cm^{2}\,\rm H^{-1}$ is the dust cross section per H at $10\,\rm K$ and Solar neighborhood dust abundance, $Z_{\rm d}$ is the normalized dust mass fraction, and $\hat{\sigma}_{\rm d}$ contains any variations in the intrinsic grain properties relative to the fiducial model. Dust properties in dense prestellar cores are somewhat uncertain and may vary considerably, so we will retain arbitrary values in our expressions. Our fiducial dust model will be $Z_{\rm d}=\hat{\sigma}_{\rm d} = 1$ and $\beta=1.5$, corresponding to model ``\texttt{a}'' used by \citetalias{masunaga1998}, based on a fit to the results of \citet{1994ApJ...421..615P}. It should be noted that this dust model lies at the low end of the range of opacity values at $\sim 10\,\rm K$  compared to various other dust models in common use \citep[e.g.][]{draine.weingartner.2001,semenov_2003}, but on the other hand dust models calibrated to diffuse ISM diagnostics may not represent the conditions in dense cores where significant grain growth may have occurred. In general, $A_{\rm d}$ and $\beta$ must be adjusted here to determine specific prediction of a given dust model.

\subsection{Optically-thin cooling regime}
\label{sec:thin}
\begin{figure*}
    \includegraphics[width=\textwidth]{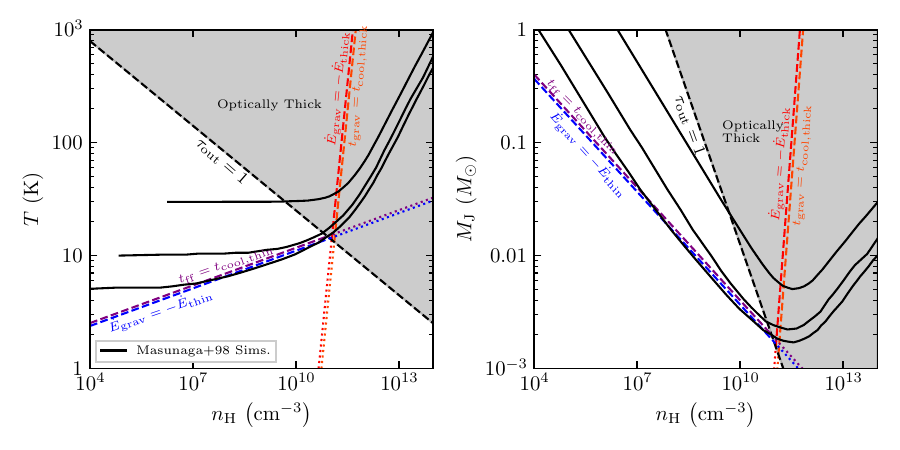}
    \caption{Map of the parameter space of density versus temperature (left) or Jeans mass $M_{\rm J}$ (right) for collapsing cores cooled by dust emission. Solid curves plot the various simulation results for the central temperature in the \citetalias{masunaga1998} radiation hydrodynamics simulations, using their fiducial dust opacity model while varying the initial radiative equilibrium temperature from $5-30\,\rm K$, and corresponding black dotted lines correspond to the prediction of the model given in \S\ref{sec:physics}. Dashed/dotted lines plot various thresholds, as labeled, corresponding to Eqs. \ref{eq:T_ff}, \ref{eq:tcool_tff}, \ref{eq:ntau}. The shaded regions are optically-thick ($\tau \gtrsim 1$), assuming our fiducial dust opacity model (Eq. \ref{eq:sigmadust}) and a Jeans-scale clump (Eq. \ref{eq:tau}).}
    \label{fig:n_vs_T}
\end{figure*}

Thermal emission by dust will be optically-thin if the optical depth $\tau_{\rm out}$ to the emitted radiation is less than unity. For the central core of the collapsing region, we make the assumption of \citetalias{lowlyndenbell1976} that the shielding length is on the order of the Jeans length , and hence
\begin{equation}
    \tau_{\rm out} \approx f_{\rm J} \lambda_{\rm J} \sigma_{\rm d} n_{\rm H} = 0.25 Z_{\rm d} \hat{\sigma}_{\rm d}\left(\frac{n_{\rm H}}{10^{10}\,\mathrm{cm}^{-3}}\right)^{1/2} \left(\frac{T_{\rm d}}{10\,\rm K}\right)^\frac{2\beta+1}{2},
    \label{eq:tau}
\end{equation}
 For a general radiation field where the radiation and dust temperatures are distinct, it is important to distinguish between $\tau_{\rm out}$ and the optical depth to ingoing radiation $\tau_{\rm in}$, given by substituting $T_{\rm rad}$ for $T_{\rm d}$ in the above expression. While $\tau_{\rm out} << 1$, the dust cooling rate per H nucleus may be written
\begin{equation}
    \dot{E}_\mathrm{thin}= -a c \sigma_{\rm d} T_{\rm d}^4
    \label{eq:lambda_thin}
\end{equation} 
where $a$ is the radiation constant and $c$ is the speed of light. While Eq. \ref{eq:lambda_thin} holds, the gas-dust temperature follows from equating it to the dominant heat source. Assuming $T\sim T_{\rm d}$, the cooling time in the optically-thin regime is a function of temperature and dust properties only:
\begin{equation}
    t_{\rm cool,thin} = \frac{-k_{\rm B}T}{\left(\gamma - 1\right) \dot{E}_{\rm thin}}= 700\mathrm{yr}\left(\frac{T}{10\mathrm{K}}\right)^{-\beta-3} \left(Z_{\rm d} \hat{\sigma}_{\rm d}\right)^{-1}.
\end{equation}
If at any time $t_{\rm cool} > t_{\rm ff}$, the fragment will virialize and quasi-isothermal collapse cannot be maintained. The ratio of the two timescales is 
\begin{equation}
    \frac{t_{\rm cool,thin}}{t_{\rm ff}}=1.6 \left(\frac{T}{10\mathrm{K}}\right)^{-\beta-3} \left(Z_{\rm d} \hat{\sigma}_{\rm d}\right)^{-1} \left(\frac{n_{\rm H}}{10^{10}\,\rm cm^{-3}}\right)^{1/2}.
    \label{eq:tcool_tff}
\end{equation}

Compressional heating  (i.e. ``$P\mathrm{d}V$ work'') due to gravitational collapse is the heat source that is always present in this scenario. The mechanical heating rate per H nucleus of gas collapsing on the free-fall timescale $\sim \left(G \rho\right)^{-1/2}$ may be written as \citepalias{lowlyndenbell1976,masunaga1998}:
\begin{equation}
        \dot{E}_{\rm grav} =C_{\rm grav} \sqrt{4 \uppi G \rho} \frac{c_{\rm s}^2 m_{\rm p}}{X_{\rm H}} \\   
    \label{eq:gravheating}
\end{equation}
where $C_{\rm grav}$ encodes the specific rate of collapse and \citetalias{masunaga1998} calibrated $C_{\rm grav}\approx 1$ from their simulations. Note that the condition $\dot{E}_{\rm grav}=-\dot{E}_{\rm cooling}$ and the condition $t_{\rm cool} \sim t_{\rm ff}$ are identical, modulo order-unity factors in the different quantities' definitions (see Fig. \ref{fig:n_vs_T}). So a collapsing clump heated only by $P\mathrm{d}V$ work will always trace a track in the $n_{\rm H}-T$ plane corresponding to $t_{\rm cool}/t_{\rm ff} \sim 1$.

Equating Eq. \ref{eq:gravheating} to the optically-thin dust cooling rate gives the dust and gas temperature where compression heating balances dust cooling:
\begin{equation}
\begin{split}
    T_{\rm grav} &= \left(\frac{2 C_{\rm grav}}{a c A_{\rm d} \mu} \sqrt{\frac{\uppi G m_{\rm p} n_{\rm H}}{X_{\rm H}^3}}\right)^\frac{1}{3+\beta}\\
    &\approx 10\mathrm{K} \, \left(Z_{\rm d} \hat{\sigma}_{\rm d}\right)^{-\frac{1}{3+\beta}}\left(\frac{n_{\rm H}}{10^{10} \rm cm^{-3}}\right)^\frac{1}{2\beta + 6}.
    \label{eq:T_ff}
\end{split}
\end{equation}
Hence, in the absence of heat sources other than compression, optically-thin cores do not collapse isothermally: they will rise gradually in temperature as they collapse, which can be seen in the lowest-temperature calculations in radiation hydrodynamics simulations \citep{masunaga1998,vaytet_2017_protostellar_collapse}.

If the absorption of radiation by dust is the main heat source and the core is optically-thin to both ingoing and outgoing radiation, then the gas and dust will approach an equilibrium temperature that is independent of the dust abundance and the normalization of the opacity, because emission and absorption are both $\propto A_{\rm d}$. This temperature,
\begin{equation}
        T_\mathrm{abs} = \left(\frac{T_{\rm rad}^\beta u_\mathrm{rad}}{a}\right)^\frac{1}{4+\beta} \approx 6  \mathrm{K} \left(\frac{T_{\rm rad}}{20 \,\rm K}\right)^\frac{1}{3} \left(\frac{u_{\rm rad}}{1 \,\rm eV\,\mathrm{cm}^{-3}}\right)^\frac{1}{6},
    \label{eq:TIR}
\end{equation}
depends only on $\beta$, the local density of radiation $u_{\rm rad}$, and spectral energy distribution represented by the effective radiation temperature $T_{\rm rad}$. This will typically set the minimum temperature in well-shielded regions of giant molecular clouds. The radiation may come from the surrounding galaxy, nearby newly-formed stars, or the cosmic microwave background. For black-body radiation, $u_{\rm rad}=aT_{\rm rad}^4$, so $T_{\rm abs}=T_{\rm rad}$, and e.g. for the CMB $T_{\rm CMB}=2.73 \mathrm{K}\left(1+z\right)$, which may be important for star formation at high redshift (see \S\ref{sec:discussion:imf}).

If $T_{\rm rad} > T_{\rm d}$ then the core will become optically-thick to the background radiation before it becomes optically-thick to its own emission. This causes a transition to a new regime. Instead of reaching the central region, incident radiation will be absorbed at larger radii and re-radiated at longer wavelengths (lower $T_{\rm rad}$), with a limiting optically-thick value
\begin{equation}
    T_{\rm bb} = \left(\frac{u_{\rm rad}}{a}\right)^{1/4} \approx 4\,\mathrm{K}\left(\frac{u_{\rm rad}}{1\,\mathrm{eV\,cm^{-3}}}\right)^{1/4}.
    \label{eq:Tbb}
\end{equation}
Strictly speaking this limit corresponds to radiative equilibrium where both $\tau_{\rm in}>>1$ and $\tau_{\rm out}>>1$, and of course intermediate temperatures will be reached e.g. if $\tau_{\rm in}>>1$ but $\tau_{\rm out} <<1$: dust re-radiated from a thin absorption zone at $T_{\rm abs}$ will reach the center with density equal to $u_{\rm rad}$. But this marginal case would give a temperature only slightly above Eq. \ref{eq:Tbb} for plausible values of $T_{\rm rad}$ and $u_{\rm rad}$. The net result is that both the radiation and gas/dust temperatures at the center will drop from $T_{\rm rad}$ and $T_{\rm abs}$ respectively, to $ \sim T_{\rm bb}=\left(u_{\rm rad}/a\right)^{1/4}$ as $\tau_{\rm in}$ crosses unity. For the purposes of our present calculation, this simply means that cases where $T_{\rm rad} > T_{\rm bb}$ initially will map onto cases where the original incident radiation field was already a black-body at $T_{\rm bb}$; in effect, the case where $T_{\rm rad} > T_{\rm bb}$ can effectively be ignored because this condition cannot persist all the way through the optically-thick and adiabatic transition. An isotropic ambient radiation field with $T_{\rm rad} < T_{\rm abs}$ (i.e. $u_{\rm rad} > a T_{\rm rad}^4$) is also not thermodynamically possible to create, so although we have assumed an {\it a priori} general radiation field, this does not actually probe any new parameter space for the opacity limit. The parameter space for the opacity limit reduces to just $u_{\rm rad}$ and the dust opacity law.

\subsubsection{General optically-thin behavior}
As long as the system is optically-thin to cooling radiation and our other assumptions hold, $T$ may be approximated by equating compressional plus radiative heating with radiative cooling, which can only be solved numerically in general. However we may approximate the overall evolution interpolating between the temperatures set by the dominant heat source, and accounting for the case where the clump is optically-thick to the ambient radiation field:
\begin{equation}
    T_{\rm thin} \approx \begin{cases}
    \max\left(T_{\rm grav},T_{\rm abs}\right)  & \tau_{\rm in} < 1\\
    \max\left(T_{\rm grav},T_{\rm bb}\right) & \tau_{\rm in} \geq 1
    \end{cases}
\end{equation}
This behaviour is seen in the simulation starting at $5 \rm K$ in \citetalias{masunaga1998} plotted in Fig. \ref{fig:n_vs_T}: it starts out in the radiation-domianted regime with $T\sim T_{\rm bb} \sim const.$, but eventually the temperature starts to rise as we transition to compression-heated regime $T \sim T_{\rm grav} \propto n^{\frac{1}{2\beta+6}}$. Note that in any of the above regimes for optically-thin cooling, the Jeans mass will continue to drop for any plausible value of $\beta$ -- the adiabatic transition cannot occur until $\tau_{\rm out} \gtrsim 1$. 

\subsection{Optically-thick cooling regime}
At a given temperature $T$, the clump will be optically-thick to its own emission if $\tau_{\rm out} > 1$, at a density threshold of 

\begin{equation}
\begin{split}
    n_{\rm thick}&= \frac{G \mu m_{\rm p}^2  T^{-2\beta - 1} }{\uppi A_{\rm d} f_{\rm J}^2 k_{\rm b} X_{\rm H}} \\
    &\approx 1.5\times 10^{11} \,\mathrm{cm}^{-3}\,\left(Z_{\rm d}\hat{\sigma}_{\rm d}\right)^{-2}\left(\frac{T}{10\,\rm K}\right)^{-2\beta-1},
    \label{eq:ntau}
\end{split}
\end{equation}
which we plot as a dashed line in the $n_{\rm H}-T$ plane in Fig. \ref{fig:n_vs_T}. Above $n_{\rm thick}$, the maximum total cooling rate per H due to optically-thick transport of dust-emitted photons is approximated asymptotically by the radiative diffusion rate \citep{masunaga:opacity.limit.tau,rafikov2007}:
\begin{equation}
    \dot{E}_{\rm thick} \approx -\frac{a c \sigma_{\rm d} T^4}{\tau^2} = \frac{\dot{E}_{\rm thin}}{\tau^2}.
    \label{eq:lambdathick}
\end{equation}
When optically-thick, the clump develops a photosphere in radiative equilibrium with the environment at $T_{\rm bb}$; if radiation were the only heat source then collapse could continue in radiative equilibrium indefinitely. This is evident in the simulation with $T_{\rm abs}=30\,\rm K$ shown in Fig. \ref{fig:n_vs_T}: the transition across $\tau=1$ has no significant effect on the thermal behaviour. Hence the condition $\tau \gtrsim 1$ is not sufficient to commence adiabatic evolution.

\subsection{Adiabatic transition and minimum Jeans mass}
The adiabatic transition occurs when $\dot{E}_{\rm grav}=-\dot{E}_{\rm thick}$, or equivalently when the optically-thick cooling time is longer than the freefall time \citep{masunaga1998}. Past this point, heat from gravitational collapse cannot escape as quickly as it is generated. This criterion is both necessary {\it and} sufficient to cause adiabatic evolution. The demarcation in the $n_{\rm H}-T$ plane is at a density of:
\begin{equation}
\begin{split}
    n_{\rm H}&\left(\dot{E}_{\rm thick}=-\dot{E}_{\rm grav}\right)\\
    &=5\times 10^{10}\,\mathrm{cm}^{-3} \left(\hat{\sigma}_{\rm d}Z_{\rm d}\right)^{-2/3} \left(\frac{T}{10 \rm \,K}\right)^\frac{2\beta - 4}{3},
    \label{eq:nthickff}
\end{split}
\end{equation}
which looks like a near-vertical line in Fig. \ref{fig:n_vs_T} for our fiducial $\beta=1.5$ model, and is exactly vertical for $\beta=2$. Considering heating by compression and radiation respectively, we consider 2 distinct cases for the adiabatic transition. The true adiabatic transition density will be roughly lesser of the two, while the minimum Jeans mass will be the larger.

\subsubsection{Case 1: Compression heating}
\label{sec:compression_mj}
If compression is the sole heat source, there is only one characteristic density $n_{\rm ad}$ at which $\tau \sim 1$, $t_{\rm cool} \sim t_{\rm ff}$, and $\dot{E}_{\rm grav} \sim -\dot{E}_{\rm thick} \sim -\dot{E}_{\rm thin}$ simultaneously, at the intersection of the lines in Fig. \ref{fig:n_vs_T}. This density is:
\begin{equation}
\begin{split}
    n_{\rm ad,grav} &= \left[\left(\frac{a c}{2C_{\rm grav}}\right)^{4\beta+2}\left(\frac{\mu}{k_{\rm B}}\right)^{6\beta+8} \times \right. \\ 
    &\left. \frac{ G^5 m_{\rm p}^{2 \beta+11} X_{\rm H}^{4 \beta - 3}}{f_{\rm J}^{4\beta+12} A_{\rm d}^{10}}\right]^\frac{1}{4\beta+7}\\
    &\approx  5.7\times 10^{10}\,\rm cm^{-3}  \left(\hat{\sigma}_{\rm d}Z_{\rm d}\right)^{-\frac{10}{4\beta+7}}.\\
\end{split}
\end{equation}
Extrapolating Eq. \ref{eq:T_ff} to $n_{\rm ad}$, the temperature at the adiabatic transition is
\begin{equation}
\begin{split}
    T_{\rm ad, grav}  &= \left(\frac{4 C_{\rm grav}^2 G^2 k_{\rm B} m_{\rm p}^3}{a^2 c^2 A_{\rm d}^4 f_{\rm J}^2 X_{\rm H}^4 \mu}\right)^\frac{1}{4\beta+7}\\
    &\approx 13\,\mathrm{K}\,\left(\hat{\sigma}_{\rm d}Z_{\rm d}\right)^{-\frac{4}{4\beta+7}},
\end{split}
\end{equation}
and the minimum Jeans mass is Eq. \ref{eq:jeansmass} evaluated at $n_{\rm ad, grav}$ and $T_{\rm ad, grav}$:
\begin{equation}
\begin{split}
    M_{\rm J,grav}^{\rm min} &= \left[\left(\frac{k_{\rm B}}{\mu}\right)^{9\beta+16} 
    \left(\frac{C_{\rm grav}}{a c}\right)^{2\beta+4} \times \right.\\
    &\left. \frac{\uppi^{12\beta+21} f_{\rm J}^{2\beta+3}}{2^{2\beta+3}3^{4\beta+7} G^{6\beta+10}m_{\rm p}^{9\beta+15} A_{\rm d} X_{\rm H}}\right]^\frac{1}{4\beta+7} \\
    &\approx 2.3\times 10^{-3} M_\odot \left(\hat{\sigma}_{\rm d}Z_{\rm d}\right)^{-\frac{1}{4 \beta + 7}}.
\end{split}
\label{eq:compression_mj}
\end{equation}
These expressions correspond to Eqs. 9-11 in \citet{lowlyndenbell1976}, and Eq. \ref{eq:compression_mj} agrees well with the minimum Jeans mass seen in the lowest-temperature simulation plotted in Fig. \ref{fig:n_vs_T}. Compression heating is always present, so this $M_{\rm J}^{\rm min}$ is effectively a lower bound on the dust opacity-limited minimum Jeans mass, and is indeed remarkably insensitive to dust properties: $\propto \left(\hat{\sigma}_{\rm d}Z_{\rm d}\right)^{-\frac{1}{11}}- \left(\hat{\sigma}_{\rm d}Z_{\rm d}\right)^{-\frac{1}{15}}$ for $\beta\sim 1-2$. However the temperature and density of the adiabatic transition point may both be significantly higher if dust opacity is low.

\subsubsection{Case 2: Radiation heating}
If radiative equilibrium is setting the temperature when $\dot{E}_{\rm grav}=-\dot{E}_{\rm thick}$ is crossed, then the temperature at the adiabatic transition will be $T_{\rm ad} \sim T_{\rm bb} = \left(u_{\rm rad}/a\right)^{1/4}$. The density threshold can be obtained by inserting Eq. \ref{eq:TIR} into Eq. \ref{eq:nthickff}:
\begin{equation}
\begin{split}
    n_{\rm ad, rad} &=   \left(\frac{a^2 c^2 G m_{\rm p}^3 T_{\rm bb}^{4 - 2\beta} \mu^4 X_{\rm H}}{4 \uppi^3 A_{\rm d}^2 C_{\rm grav}^2 f_{\rm J}^4 k_{\rm B}^4} \right)^{1/3} \\ 
    &\approx 5.2\times 10^{10}\,\mathrm{cm}^{-3} \left(\frac{T_{\rm ad}}{10\,\mathrm{K}}\right)^\frac{4-2\beta}{3}\left(\hat{\sigma}_{\rm d}Z_{\rm d}\right)^{-2/3},\\
\end{split}
\end{equation}
and evaluating Eq. \ref{eq:jeansmass} at $n_{\rm ad}$ and $T_{\rm ad}$ gives:
\begin{equation}
\begin{split}
    M_{\rm J,rad}^{\rm min} &= \left(\frac{\uppi^{18} A_{\rm d}^2 C_{\rm grav}^2 f_{\rm J}^4 k_{\rm B}^{13} T_{\rm bb}^{2\beta+5} X_{\rm H}^2}{11664 a^2 c^2 G^{10} m_{\rm p}^{15} \mu^{13}}\right)^{1/6}\\
    &\approx 1.7 \times 10^{-3} M_\odot \left(\frac{T_{\rm bb}}{10 \rm K}\right)^\frac{2\beta+5}{6}\left(\hat{\sigma}_{\rm d}Z_{\rm d}\right)^{1/3},
    \label{eq:mj_rad}
\end{split}
\end{equation}
which corresponds to Eq. 26 in \citetalias{masunaga:opacity.limit.tau}. For the case $T_{\rm bb}=30\,\mathrm{K}$ simulated in \citetalias{masunaga1998}, this expression predicts $M_{\rm J}^{\rm min}=7.4\times 10^{-3}M_\odot$, again in good agreement with the data in Fig. \ref{fig:n_vs_T}. If the radiation field is the cosmic microwave background with $T_{\rm bb}=2.73\left(1+z\right)\,\rm K$, then $M_{\rm J}^{\rm min} = 3\times 10^{-4}M_\odot \left(1+z\right)^\frac{2\beta + 5}{6}\left(\hat{\sigma}_{\rm d}Z_{\rm d}\right)^{1/3}$, which would exceed the compression-dominated value above $z \sim 3$, assuming $\hat{\sigma}_{\rm d}Z_{\rm d}\sim 1$.

The relevant regime  is that which results in the higher temperature, so the overall minimum Jeans mass will be roughly the larger of Eq. \ref{eq:compression_mj} and Eq. \ref{eq:mj_rad}:
\begin{equation}
    M_{\rm J}^{\rm min} \approx \max \left(M_{\rm J,rad}^{\rm min},M_{\rm J,rad}^{\rm min}\right),
\end{equation}
which we plot in Fig. \ref{fig:mj_variation}.

\subsection{Hydrostatic core mass}
\label{sec:larsonmass}
Another important mass scale that reached the first hydrostatic core before $\rm H_2$ dissociation causes the second collapse at $T_{\rm diss} \sim 1500 \rm K$ \citep{larson_1969}, the ``Larson mass'' $M_{\rm L}$. \citet{Hennebelle_2019_fhsc_tidal} relate $M_{\rm J}^{\rm min}$ to $M_{\rm L}$ by assuming the entropy of the core is determined at the opacity limit by $n_{\rm ad}$ and $T_{\rm ad}$, and the subsequent evolution is isentropic with adiabatic index $\gamma_1=5/3$ below $T_{\rm ex}=150\rm K$, and $\gamma_2=7/5$ above it. The adiabatic scaling of the Jeans mass gives:
\begin{equation}
    M_{\rm L} = k_{\rm L} M_{\rm J}^{\rm min} \left(\frac{T_{\rm ex}}{T_{\rm ad}}\right)^\frac{3\gamma_1 - 4}{2\gamma_1 -2 }\left(\frac{T_{\rm diss}}{T_{\rm ex}}\right)^\frac{3\gamma_2 - 4}{2\gamma_2 -2 } 
\end{equation}
where $k_{\rm L}$ is a dimensionless factor to account for the new hydrostatic structure after the adiabatic transition, and $k_{\rm L}\approx 1.3$ recovers the $\sim 0.03 M_\odot$ hydrostatic core mass measured in numerical simulations in the compression-heated limit \citep{vaytet_2017_protostellar_collapse}. In the compression regime we get
\begin{equation}
    M_{\rm L} = 0.03\,M_\odot \left(\hat{\sigma}_{\rm d}Z_{\rm d}\right)^\frac{2}{4 \beta + 7},
\end{equation}
which is an increasing, but still fairly-shallow function of dust opacity. In the radiative limit we have
\begin{equation}
    M_{\rm L} = 0.03 M_\odot\,\left(\frac{T_{\rm bb}}{10\,\mathrm{K}}\right)^\frac{4\beta+1}{12} \left(\hat{\sigma}_{\rm d}Z_{\rm d}\right)^\frac{1}{3} ,
    \label{eq:ML_rad}
\end{equation}
which scales in the same way with dust opacity as the corresponding Jeans mass, but is somewhat less sensitive to the ambient radiation field.

\section{Discussion}
\subsection{Comparison with  \citetalias{lowlyndenbell1976} and \citetalias{masunaga:opacity.limit.tau}}
\label{comparison}
In the previous section we derived approximate expressions for the adiabatic transition density and minimum Jeans mass under similar assumptions to \citetalias{masunaga:opacity.limit.tau}, but with one key difference that strongly affects how our predictions generalize across parameter space. 
When deriving the minimum Jeans mass, \citetalias{masunaga:opacity.limit.tau} assumed that the condition $\dot{E}_{\rm grav} \sim -\dot{E}_{\rm thin}$ was sufficient for the Jeans mass to reach a minimum (their ``Case 1''). This is not a realistic assumption. From Eqs. \ref{eq:T_ff} and \ref{eq:TIR} for the optically-thin temperature it is clear that  $T$ cannot scale as steeply as $\propto \rho^{1/3}$ for any plausible dust model $\beta \sim 1-2$. Hence the Jeans mass $M_{\rm J} \propto T^{3/2}\rho^{-1/2}$ will always decrease while the clump is optically-thin. The gas must become optically-thick to undergo an adiabatic transition or reach a minimum Jeans mass. Although \citetalias{masunaga:opacity.limit.tau} used the \citetalias{masunaga1998} simulations to guide their model, crucially there was no example of a simulation that reached its minimum Jeans mass when $\tau < 1$, and we are not aware of such a numerical calculation in the literature. We do concur that the condition $\tau \gtrsim 1$ is not {\it sufficient} for the adiabatic transition, as was assumed by \citetalias{lowlyndenbell1976}, but it is a {\it necessary} condition.

The correct choice is to ignore Case 1 and consider only the criterion $\dot{E}_{\rm grav} \approx -\dot{E}_{\rm thick}$ (their ``Case 3''), which is necessary and sufficient for the adiabatic transition. This eliminates the predicted regime in which $M_{\rm J}^{\rm min}\propto \left(Z_{\rm d}\hat{\sigma}_{\rm d}\right)^{-1}$ that would imply strong variations in the opacity limit with dust abundance or properties. A Jeans-scale core with $M_{\rm J}^{\rm min}=3.7 M_\odot$, $Z_{\rm d}\hat{\sigma}_{\rm d}=10^{-2}$, and $T=10\rm \,K$ would have an optical depth $\tau \sim 10^{-6}-10^{-5}$, and thus would be very well described by the optically-thin behavior in \S\ref{sec:thin}, and would continue to decrease in Jeans mass. Two regimes remain: the compression-dominated regime (\S\ref{sec:compression_mj}) where $M_{\rm J}$ is nearly independent of dust properties (Eq. \ref{eq:compression_mj}), and the radiation-dominated regime where $M_{\rm J} \propto \left(Z_{\rm d}\hat{\sigma}_{\rm d}\right)^{1/3}$ (Eq. \ref{eq:mj_rad}), a much milder sensitivity to dust properties.


\subsection{Implications for the IMF}
\label{sec:discussion:imf}
\begin{figure}
    \centering
    \includegraphics[width=\columnwidth]{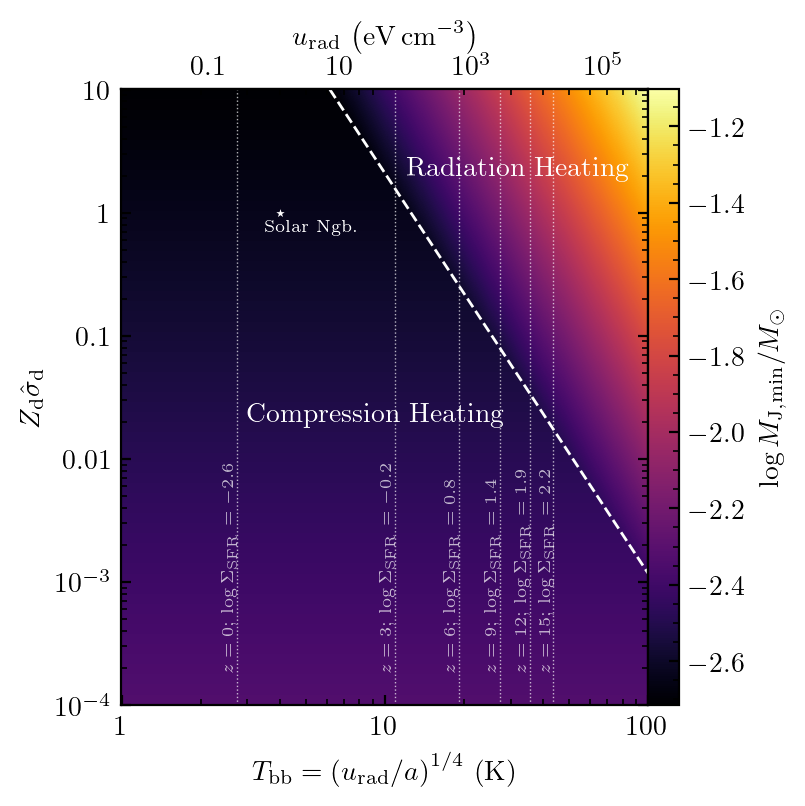}
    \caption{Variation of the minimum Jeans mass $M_{\rm J}^{\rm min}$ as a function of the Solar-normalized dust opacity parameter $Z_{\rm d}\hat{\sigma}_{\rm d}$ and the radiative equilibrium temperature $T_{\rm bb}$ (Eq. \ref{eq:Tbb}). We assume a $\beta=1.5$ dust opacity model and estimate $M_{\rm J}^{\rm min}$ as the greater of the compression-dominated limit (Eq. \ref{eq:compression_mj}) and the radiation-dominated limit (Eq. \ref{eq:mj_rad}). The dotted line demarcates the transition between the two regimes. The location of the Solar neighborhood in this parameter space is indicated, assuming $u_{\rm rad} \sim 1\,\mathrm{eV}\,\mathrm{cm}^{-3}$. Vertical lines plot the lower bound $T_{\rm bb} \geq 2.73 \left(1+z\right)\mathrm{K}$ imposed by the CMB at various indicated redshifts. Corresponding values of $\log \Sigma_{\rm SFR}$ in $M_\odot\,\mathrm{yr}^{-1}\mathrm{kpc}^{-2}$ are also indicated, assuming $u_{\rm rad} \propto \Sigma_{\rm SFR}$ and scaling from the Solar neighborhood value.}
    \label{fig:mj_variation}
\end{figure}
The opacity-limited minimum Jeans mass has long been thought to be responsible for some feature of the IMF. Most naturally, it could shape the low-mass cutoff and the abundance of brown dwarfs. \citet{bate:2005.imf.metallicity} performed controlled numerical experiments in which their varied it explicitly with a barotropic equation of state, and found that it determined the low-mass IMF cutoff, while the mean cloud Jeans mass determined the mean stellar mass \citep[e.g.][]{bate.bonnell:2005.jeansmass}. More recently, \citet{Lee_Hennebelle_2018_IC,Lee_Hennebelle_2018_EOS,Lee_Hennebelle_2019_T_B,Hennebelle_2019_fhsc_tidal,Colman_Teyssier_2019_tidal_screening} used simulations and analytic theory to explore how $M_{\rm J}^{\rm min}$ might set the IMF {\it peak}, on the order of $0.1 M_\odot$. The jump from Jupiter-like to M dwarf masses has two steps. First, from the opacity limit to the hydrostatic core $\sim 10$ times more massive (\S\ref{sec:larsonmass}). Then, there is a region around this core in which fragmentation is suppressed due to the core's tidal field, so the gas reservoir for a typical star is several times larger than the hydrostatic core mass.

In ambient radiation fields similar to the Solar neighborhood ($\sim 1 \rm \,eV\,cm^{-3}$), $M_{\rm J}^{\rm min}$ is nearly constant, scaling $\propto A_{\rm d}^{-1/13}-A_{\rm d}^{-1/15}$, leading to factor of $\sim 2$ variations over the entire parameter space where a dust-coupled phase exists. So any IMF feature set by $M_{\rm J}^{\rm min}$ should not vary strongly in galactic regions with moderate radiation fields. On the other hand, where radiation fields are strong, such as in galactic centers, starburst galaxies, or in the high-redshift Universe where the CMB is $\left(1+z\right)^4$ times more intense, Eq. \ref{eq:mj_rad} predicts non-negligible variations in $M_{\rm J}^{\rm min}$, and by extension the IMF, as a function of the ambient radiation field and metallicity. In Figure \ref{fig:n_vs_T} we map out the parameter space for variations in $M_{\rm J}^{\rm min}$ as a function of $Z_{\rm d}\hat{\sigma}_{\rm d}$ and $u_{\rm rad}$/$T_{\rm bb}$. The most severely-affected regions would be those with $Z_{\rm d}\hat{\sigma}_{\rm d} \gtrsim 1$ and $T_{\rm abs} \gtrsim 10 \rm \,K$, corresponding to regions of galaxies of Milky Way mass or greater, with ambient radiation fields $\gtrsim 100 \times$ the Solar neighborhood. Such regions would generally have a significantly less bottom-heavy IMF, or a greater IMF peak mass.

\citet{bate:2023.hiz.sf} presented a suite of star formation simulations with a $z\sim 5$ CMB covering a range of metallicities from $0.01-1$, and found their results only deviated significantly from their $z\sim 0$ results when $Z_{\rm d}\hat{\sigma}_{\rm d} \approx 1$, which produced an IMF scaled by a factor of $\sim 3$ in mass. Their dust model (detailed in \citealt{bate:2014.imf.metallicity}) is approximated by $\beta \sim 2.2$, and for these parameters our model predicts $M_{\rm J}^{\rm min}=5\times 10^{-3}$. Meanwhile the lower-metallicity $z=5$ models and \citet{bate:2014.imf.metallicity} $z=0$ models would all still be in the compression-dominated regime with $M_{\rm J}^{\rm min}\sim 2\times 10^{-3}M_\odot$. Therefore the results of these simulations are fairly well-fit by model in which the IMF is determined by the opacity limit.

Once star formation has begun in a cluster, the ambient radiation field may be of secondary importance to that generated by the star cluster itself, particularly once the spatial correlation of young stars is taken into account \citep{lee.hopkins:2020.sf.radfield}. For example, in the Solar neighborhood GMC simulated in \citet{grudic:2022.starforge.full.physics} we found that the radiation field seen by a median H atom in the cloud was elevated by a factor of 100 compared to the ISRF once massive stars started appearing - and the enhancement in dense regions forming most of the stars would be significantly greater. In this scenario too the opacity limit would move to higher masses and the formation of very low-mass stars would be suppressed, although it would have been only marginally-resolved at the $10^{-3}M_\odot$ mass resolution of our simulation. Note that this scenario, in which protostellar radiation modifies the opacity limit, is distinct from the {\it local} self-regulation of the Jeans mass around individual stars due to protostellar radiation, as explored in other works \citep{bate_2009_rad_importance, krumholz_stellar_mass_origin}.

\subsubsection{Importance of other physics for the IMF}
The IMF is probably not determined by the opacity limit alone. This was noted by \citet{bate.bonnell:2005.jeansmass}, who also found sensitivity to the cloud bulk properties in hydrodynamic simulations with a barotropic equation of state.  It is also evident from the results of the isothermal MHD star formation experiments of \citet{Haugbolle_Padoan_isot_IMF} and \citet{guszejnov_isothermal_mhd}, neither of which included dust opacity physics in any form. Both works demonstrated a turnover feature in the mass spectrum from the Salpeter-like $\mathrm{d}N/\mathrm{d}M \sim M^{-2}$ to the flat $\mathrm{d}N/\mathrm{d}M \sim M^{-1}$, argued by \citet{Haugbolle_Padoan_isot_IMF} to be related to the turbulence-compressed Jeans mass. The GMC simulations with protostellar jet feedback by \citet{starforge_jets_imf} also included an isothermal run that produced a very similar IMF to the runs with realistic cooling physics (except at small masses), indicating that those IMF results were not being determined by the opacity limit. These results suggest that the thermal structure of GMCs at densities much lower than the opacity limit can also influence the IMF. This is qualitatively consistent with the predictions of turbulent fragmentation theories \citep{padoan_nordlund_2002_imf,2008ApJ...684..395H, 2013MNRAS.430.1653H}.

Another important regulator of fragmentation is accretion-powered protostellar radiation mediated by dust heating, which inevitably dominates the thermal structure of protostellar disks, suppressing disk fragmentation and brown dwarf formation \citep{bate_2009_rad_importance,Offner_2009_radiative_sim}. The fundamental physics of protostellar evolution and radiation have even been proposed to determine the characteristic mass of the IMF more generally, if the thermal structure of individual collapsing fragments can be regulated by radiation from the central star \citep{krumholz_stellar_mass_origin,krumholz2016, guszejnov_feedback_necessity,Federrath_2017_IMF_converge_proceedings}. However, in this picture one finds that the self-regulated fragment mass decreases with increasing ambient ram pressure and/or surface density \citep{krumholz_stellar_mass_origin,tab:2022.imf.variation}. 
The detailed numerical prediction in Fig. 5 of \citet{krumholz_stellar_mass_origin} is well-approximated by
\begin{equation}
    M_{\rm K} \approx 0.03 M_\odot \left(\frac{P}{10^{10}\rm cm^{-3}}\right)^{-0.28} \left(\hat{\sigma}_{\rm d} Z_{\rm d}\right)^{-0.22}.
\end{equation}
This might explain why \citet{2020ApJ...904..194H} did not find that their IMF peak prediction was strongly affected by the addition of protostellar radiation: they simulated very dense star-forming clumps with pressures upwards of $G \Sigma^2 \sim 10^8-10^{10} \,k_{\rm B}\,\rm cm^{-3}\,\rm K$, where the Krumholz mass is predicted to be at most comparable to $M_{\rm L}$, making the effects difficult to distinguish.

The natural synthesis of these findings is that initial and boundary conditions should be taken into account when asking which processes determine the IMF. There are at least three distinct, physically-plausible fragment mass scales that could produce corresponding IMF features: one related to the opacity limit, another related to the physics of turbulent fragmentation, and another to protostellar heating. 


\section{Conclusion}

In this work we have revisited the physics of dust-cooled protostellar collapse, under a set of analytic asymptotic approximations guided and validated by numerical simulations. We have adopted a physically-consistent model in which all relevant quantities are expressed in terms of dust properties and ambient conditions, and adopted calibrations in quantitative agreement with the results of radiation hydrodynamics simulations (e.g. Fig. \ref{fig:n_vs_T}). Crucially, we have avoided fallacious assumptions about the conditions for the adiabatic transition (c.f. \S\ref{comparison}). Lastly, we have explored some implications of the environmental variation of the minimum Jeans mass in the context of the results of recent star formation simulations. Nothing we have done here is fundamentally new, but we have aimed to put a fresh coat of paint on these important concepts, and to sort out some confusing or conflicting results.

Our key result is that the opacity-limited minimum Jeans mass is roughly the larger of a compression-heated limit, Eq. \ref{eq:compression_mj} (Eq. 11 in \citealt{lowlyndenbell1976}), and a radiation-heated limit, Eq. \ref{eq:mj_rad} (generalizing Eq. 26 in \citealt{masunaga:opacity.limit.tau}). The implied variations with metallicity and the radiative equilibrium temperature (Fig. \ref{fig:mj_variation}) may warrant consideration in modern discussions of the role of the opacity limit or the Larson mass in setting the IMF's lower cutoff or peak, particularly in extreme or high-redshift conditions.

\section*{Acknowledgements}
We are grateful for helpful feedback from Stella Offner, Matthew Bate and Patrick Hennebelle. Support for MYG was provided by NASA through the NASA Hubble Fellowship grant \#HST-HF2-51479 awarded  by  the  Space  Telescope  Science  Institute,  which  is  operated  by  the   Association  of  Universities  for  Research  in  Astronomy,  Inc.,  for  NASA,  under  contract NAS5-26555. This work used computational resources provided by Frontera allocation AST21002. This research is part of the Frontera computing project at the Texas Advanced Computing Center. Frontera is made possible by National Science Foundation award OAC-1818253.





\bibliographystyle{mnras}
\bibliography{bibliography} 

\end{document}